\documentclass[twocolumn,showpacs,preprintnumbers,amssymb,epsfig,prl,aps]{revtex4}

\usepackage{graphicx}
\usepackage{dcolumn}
\usepackage{bm}

\begin{document}

\preprint{}

\title{Superconductor - Normal and Quantum Superconductor-Insulator
Transition at the LaAlO$_{3}$/SrTiO$_{3}$Interface}
\author{T. Schneider}
\affiliation{Physick-Institute, University of Zurich, Winterthurerstrasse 190, 8057 Zurich,
Switzerland. }
\email{tschnei@physik.unizh.ch}
\author{A.D. Caviglia}
\author{S. Gariglio}
\author{N. Reyren}
\author{D. Jaccard}
\author{J.-M. Triscone}
\affiliation{DPMC, University of Geneva, 24 Quai Ernest-Ansermet, 1211 Geneva 4,
Switzerland. }

\date{\today }

\begin{abstract}
Superconductivity at the interface between the insulators LaAlO$_{3}$ and
SrTiO$_{3}$ has been tuned with the electric field effect. The data provide
evidence for a two dimensional quantum superconductor to insulator (2D-QSI)
transition. Here we explore the compatibility of this phase transition line
with Berezinskii-Kosterlitz-Thouless (BKT) behavior and a 2D-QSI transition.
In an intermediate regime, limited by a finite size effect, we uncover
remarkable consistency with BKT- criticality, weak localization in the
insulating state and non-Drude behavior in the normal state. Our estimates
for the critical exponents of the 2D-QSI-transition, $z\simeq 1$ and $%
\overline{\nu }\simeq 2/3$, suggest that it belongs to the 3D-xy
universality class.
\end{abstract}

 \maketitle

The conducting interface between LaAlO$_{3}$ and SrTiO$_{3}$, two excellent
band insulators, has been attracting a lot of attention\cite%
{ohtomo,okamoto,willmott,siemons,herranz}. Recently, two different ground
states, magnetic \cite{brinkman} and superconducting \cite{reyren}, have been
experimentally identified. In a very recent report \cite{andrea}, it was
shown that the electric field effect can be used to map the phase diagram of
this interface system. As the carrier density is increased the system
undergoes a 2D-QSI transition. A further increase reveals a superconducting
dome. Moreover it was shown that the characteristics of the
superconducting transition are consistent with a superconducting sheet of
about 10 nm thick \cite{reyren2}.

 Here we attempt to unravel the nature of the phase transition line and its
endpoint which separates the superconducting from the insulating ground
state. For this purpose we explore the compatibility of the phase transition
with Berezinskii-Kosterlitz-Thouless (BKT) critical behavior \cite
{berz,kosterlitz} and of its endpoint, where superconductivity disappears,
with a two dimensional quantum superconductor to insulator (2D-QSI)
transition associated with weak localization \cite{lee}. In contrast to the
disorder tuned QSI transition, electrostatic tuning changes the carrier
density without altering the disorder landscape \cite{ahn}. In an
intermediate temperature regime we uncover remarkable agreement with
BKT-criticality in the superconductor to normal state transition, weak
localization in the insulating phase, and non-Drude behavior in the normal
state. It is shown that both conduction mechanisms are limited by a finite
size effect, whereupon the BKT-correlation length and the diverging length
associated with weak localization \cite{lee} cannot grow beyond a limiting
length, set by the extent of the homogeneous regions. Our analysis also
reveals that the electrostatic tuned BKT-phase transition line ends at a
2D-QSI critical point which appears to fall onto the universality class of
the classical 3D-xy model. In addition we explore the $T_{c}$ dependence of
the vortex core radius and the vortex energy. These properties appear to be
basic ingredients to understand the $T_{c}$ variation.

To explore the compatibility with BKT critical behavior we invoke the
characteristic temperature dependence of the correlation length above $T_{c}$
\cite{kosterlitz},
\begin{equation}
\xi =\xi _{0}\exp \left( 2\pi /\left( bt^{1/2}\right) \right) ,t=\left\vert
T/T_{c}-1\right\vert ,  \label{eq1}
\end{equation}
where $\xi _{0}$ is the classical vortex core radius and $b$ is
related to the energy needed to create a vortex \cite{ambek,finotello,steel}. Note that $b$ also enters the temperature dependence of the magnetic penetration depth $\lambda $ below the universal Nelson-Kosterlitz jump: $
\lambda ^{2}\left( T_{c}\right) /\lambda ^{2}\left( T\right) =\left(
1+b\left\vert t\right\vert ^{1/2}/4\right) $ \cite{ambek}. Moreover, $b$ is
related to the vortex energy $E_{c}$ in terms of \cite{steel,dahm}
\begin{equation}
b=4\pi T_{c}^{1/2}/b_{R}=f\left( E_{c}/\left( k_{B}T_{c}\right) \right) .
\label{eq1a}
\end{equation}
Invoking dynamic scaling the resistance $R$ scales in $D=2$ as \cite{book}
\begin{equation}
R\propto \xi ^{-z_{cl}},  \label{eq2}
\end{equation}
where $z_{cl}$ is the dynamic critical exponent of the classical dynamics.
$z_{cl}$ is usually not questioned to be anything but the value that
describes simple diffusion: $z_{cl}=2$ \cite{pierson}. Combining these
scaling forms and taking the occurrence of a 2D-QSI transition at $R_{0c}=$\
$R_{0}\left( T_{c}=0\right) $ into account we obtain%
\begin{equation}
R\left( T\right) =R_{0}\exp \left( {-b_{R}}(T-T_{c})^{-1/2}\right),
\label{eq3}
\end{equation}
with
\begin{equation}
b_{R}=4\pi T_{c}^{1/2}/b,\text{ }\Delta R_{0}=R_{0c}-R_{0}\propto 1/\xi
_{0}^{2},  \label{eq4}
\end{equation}
because $\Delta R_{0}$ is the singular part of $R_{0}$ close to the
2D-QSI transition. To assess the compatibility with the characteristic BKT-behavior we analyze
the data in terms of
\begin{equation}
\left( d\ln R/dT\right) ^{-2/3}=\left( 2/b_{R}\right) ^{2/3}(T-T_{c}).
\label{eq5}
\end{equation}

\begin{figure}[tbp]
\centering
\includegraphics[angle=0,width=8.6cm]{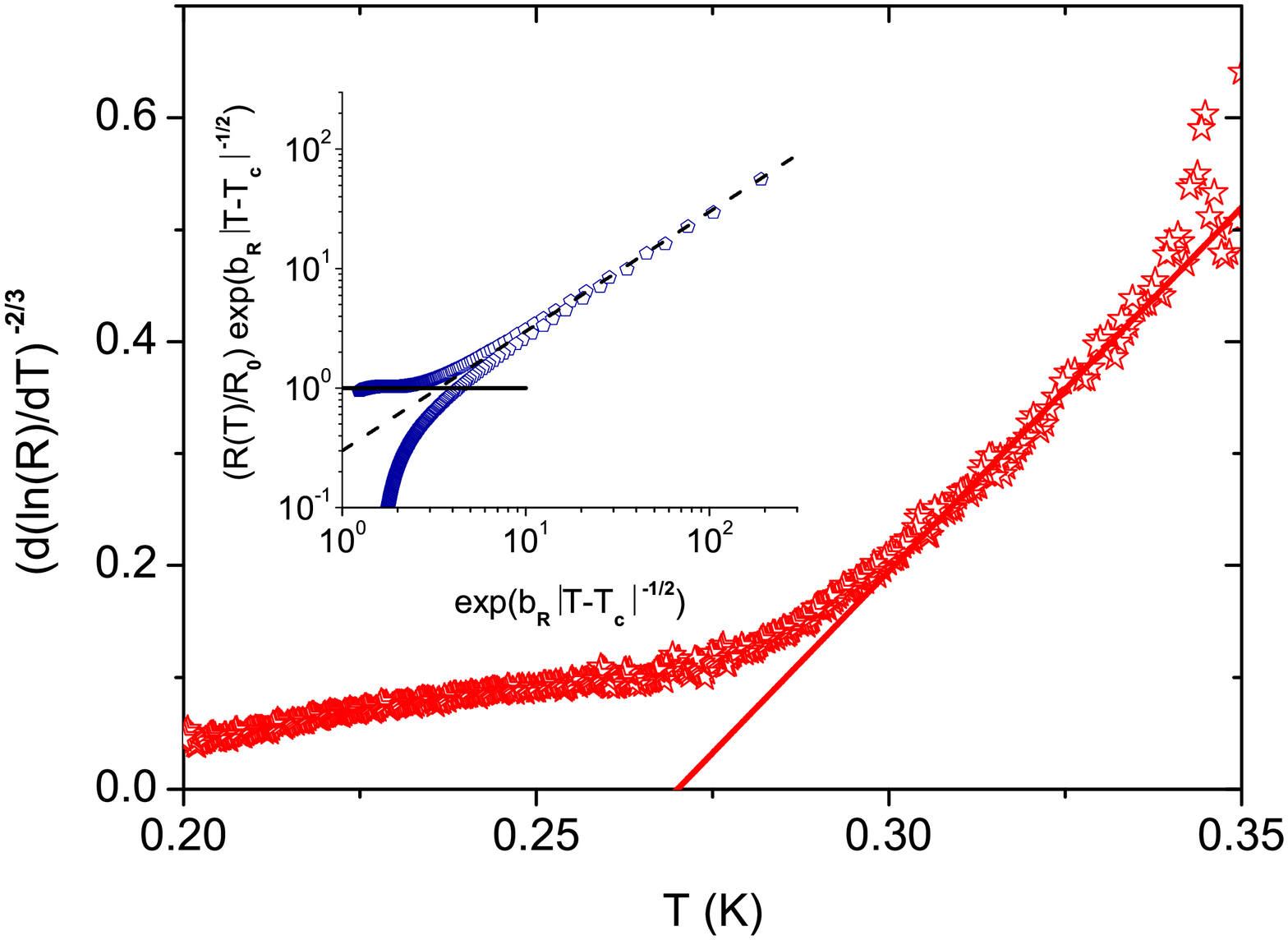}
\caption{ $\left( d\ln R/dT\right) ^{-2/3}$ \textit{vs}. $T$ for $
V_{g}=40$ V from Caviglia \textit{et al}. \cite{andrea} where $
R=3/5R_{\square }$. The solid line is $\left( d\ln R/dT\right)
^{-2/3}=6.5\left( T-T_{c}\right) $ yielding the estimates $T_{c}=0.27$ K and
$\left( 2/b_{R}\right) ^{2/3}=6.5$; the inset shows $R$ exp$\left(
b_{R}\left\vert T-T_{c}\right\vert ^{-1/2}\right) /R_{0}$ \textit{vs}. exp$
\left( b_{R}\left\vert T-T_{c}\right\vert ^{-1/2}\right) $ with $R_{0}=1670$
$\Omega $. The upper branch corresponds to $T>T_{c}$ and the lower one to $
T<Tc$. The solid line is $\left( R/R_{0}\right) $exp$\left( b_{R}\left\vert
T-T_{c}\right\vert ^{-1/2}\right) =1$ and the dashed one $\left(
R/R_{0}\right) $exp$\left( b_{R}\left\vert T-T_{c}\right\vert ^{-1/2}\right)
=0.4$exp$\left( b_{R}\left\vert T-T_{c}\right\vert ^{-1/2}\right) \propto
\left( \xi /L\right) ^{2}$.}
\label{fig1}
\end{figure}

 Fig. \ref{fig1} shows \ $\left( d\ln R/dT\right) ^{-2/3}$ \textit{vs}. $T$
for $V_{g}=40$ V.  $V_{g}$ is the gate voltage allowing to tune the carrier
density. In spite of the rounded transition there is an intermediate
regime revealing the characteristic BKT-behavior (\ref{eq5}). Noting that
real systems are homogeneous over a limited domain only, the rounding may be
attributable to a standard finite size effect, whereupon the correlation
length cannot grow beyond a limiting length, set by the lateral extent $L$
of the homogenous regions \cite{book,fisherm}. In this case $R\left(
T\right) $ scales as $R\left( T\right) $exp$\left( b_{R}\left\vert
T-T_{c}\right\vert ^{-1/2}\right) /R_{0}=f(x)$. $f(x)$ is the finite size
scaling function with $x=$exp$\left( b_{R}\left\vert T-T_{c}\right\vert
^{-1/2}\right) /L^{2}\propto \xi ^{2}/L^{2}$. If $\xi <L$ critical behavior
can be observed as long as $f(x)\simeq 1$. This regime corresponds to the
horizontal line in the inset of Fig. \ref{fig1}. If $\xi >L$ the scaling
function approaches $f(x)\propto x$ so that $R(T)/R_{0}$ tends to $L^{-2}$,
the behavior indicated by the dashed line. In this context it should be
recognized that the BKT correlation length does not exhibit the usual and
relatively slow algebraic divergence as $T_{c}$ is approached (Eq. (\ref{eq1}
)). For this reason the elimination of the finite size effect would require
unprecedented sample homogeneity. The emerging agreement with BKT behavior,
limited by a standard finite size effect, allows us to discriminate the
rounded transition from other scenarios, including strong disorder which
destroys the BKT behavior. It also provides the basis to estimate $T_{c}$, $
b_{R}\left( T_{c}\right) $ and $R_{0}\left( T_{c}\right) $, and with that $
b\left( T_{c}\right) $ and $\xi _{0}\propto \Delta R_{0}^{-1/2}$ with
reasonable accuracy.

\begin{figure}[tbp]
\centering
\includegraphics[angle=0,width=8.6cm]{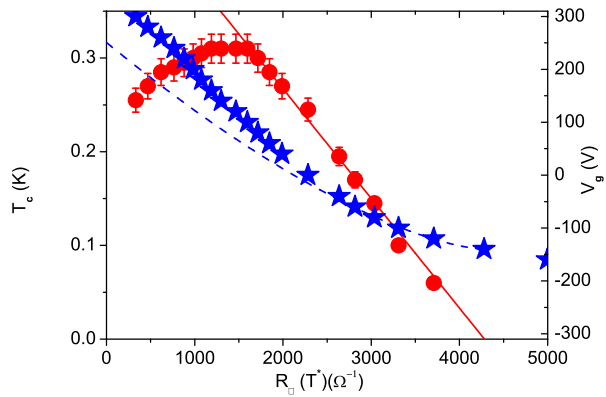}
\caption{$T_{c}$ \textit{vs}. $R_{\square }\left( T^{\ast }\right) $
$\left(\bullet \right) $and $V_{g}$ \textit{vs}. $R_{\square }\left(
T^{\ast }\right) \left(\bigstar\right) $ at $T^{\ast }=0.4$ K from Caviglia
\textit{et al}. \cite{andrea}. The solid line is $T_{c}$ $=1.17\cdot
10^{-4}\Delta R(T^{\ast }$ $)$ and the dashed one $V_{g}$ $=V_{gc}+1.39\cdot
10^{-3}$ $\Delta R^{3/2}(T^{\ast })$ with $\Delta R\left( T^{\ast }\right)
=(R_{\square c}\left( T^{\ast }\right) -R_{\square }\left( T^{\ast }\right) )
$, $R_{\square c}\left( T^{\ast }\right) =4.28$ k$\Omega $ and $V_{gc}=-140$
V.}
\label{fig2}
\end{figure}

Applying this approach to the $R\left( T\right) $ data for each gate voltage
$V_{g}$ we obtain the BKT-transition line depicted in Fig. \ref{fig2},
displayed as $T_{c}$ \textit{vs}. $R_{\square }\left( T^{\ast }\right) $,
the normal state sheet resistance at $T^{\ast }=0.4$K. We observe that it
ends around $R_{\square c}\simeq 4.28$ k$\Omega $ where the system is
expected to undergo a 2D-QSI transition because $T_{c}$ vanishes. With
reduced $R_{\square }$ the transition temperature increases and reaches its
maximum value, $T_{cm}\simeq 0.31$ K, around $R_{\square }\simeq 1.35$ k$
\Omega $. With further reduced resistance $T_{c}$ decreases. To identify
deviations from Drude behavior ($\sigma \propto n$) in the normal state,
which will be discussed later, we also included the gate voltage dependence
of the normal state resistance.

\begin{figure}[tbp]
\centering
\includegraphics[angle=0,width=8.6cm]{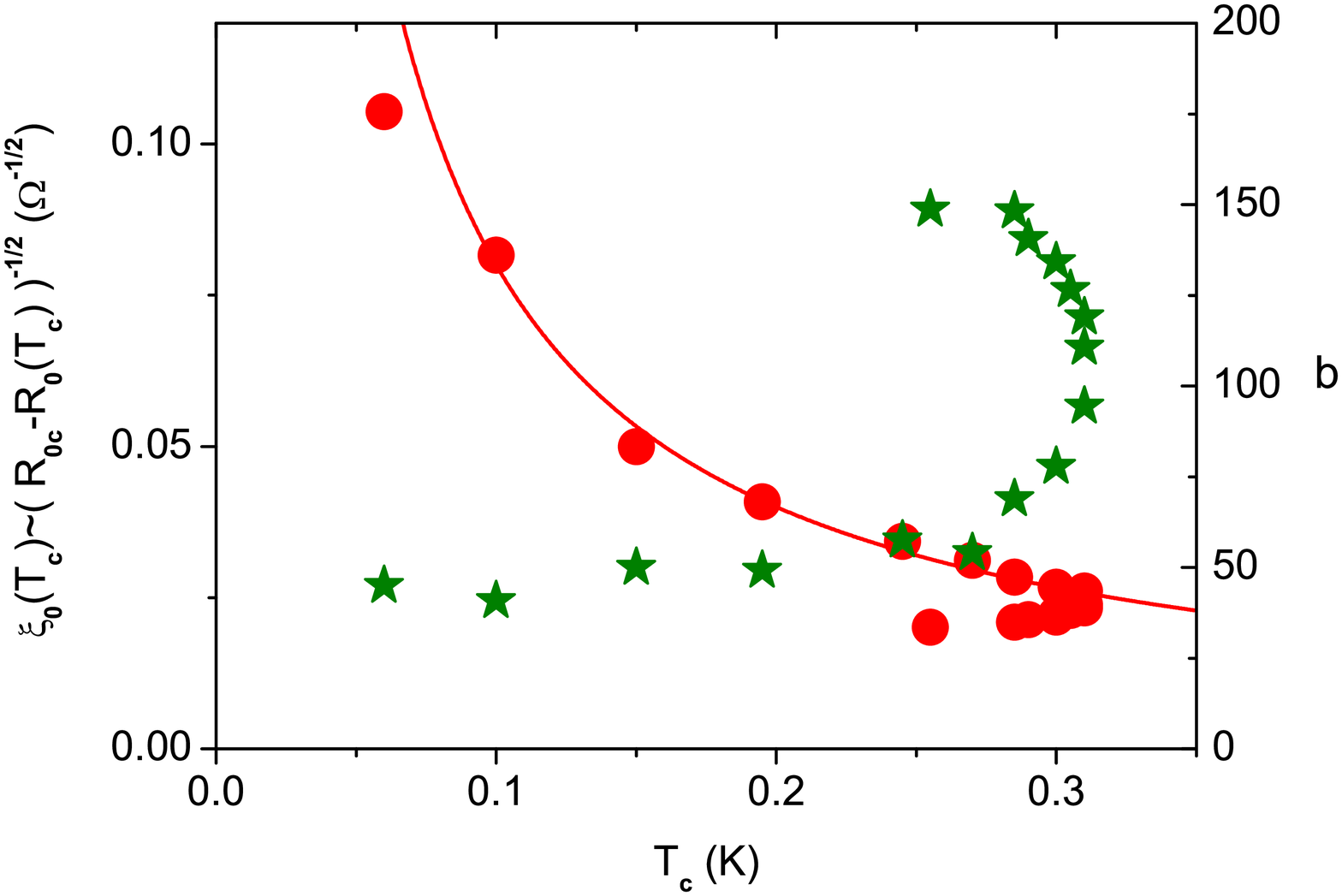}
\caption{ $\xi _{0}\left( T_{c}\right) \propto \left(
R_{0c}-R_{0}\left( T_{c}\right) \right) ^{-1/2}\left( \bullet \right) $ and $
b$ $\left( \bigstar \right) $ \textit{vs }. $T_{c}$ from Caviglia \textit{et
al}. \cite{andrea} where $R=3/5R_{\square }$. The solid line is $\xi _{0}\left(
T_{c}\right) \propto \left( R_{0c}-R_{0}\left( T_{c}\right) \right)
^{-1/2}=8\cdot 10^{-3}/T_{c}$ with $R_{0c}=2700$ $\Omega $.}
\label{fig3}
\end{figure}

 According to the scaling theory of quantum critical phenomena one expects
that close to the 2D-QSI transition $T_{c}$ scales as \cite{book,kim}
\begin{equation}
T_{c}\propto \delta ^{z\bar{\nu}},  \label{eq5a}
\end{equation}
where $\delta $ is the appropriate scaling argument, measuring the relative
distance from criticality. $\overline{\nu }$ denotes the critical exponent
of the zero temperature correlation length $\xi \left( T=0\right) \propto
\delta ^{-\overline{\nu }}$ and $z$ the dynamic critical exponent. From Fig. \ref{fig2} it is seen that the experimental data point to the relationship%
\begin{equation}
T_{c}\propto \Delta R\left( T^{\ast }\right) \propto \Delta V_{g}^{2/3},
\label{eq6}
\end{equation}
close to quantum criticality. If the scaling argument is $\Delta R\left(
T^{\ast }\right) $,\ $z\overline{\nu }=1$, while if $\delta =\Delta V_{g}$, $z\overline{\nu }=2/3$. Since the measured modulation of the charge density $\Delta n_{2D}$ induced by the gate voltage scales in the regime of interest
as \cite{andrea}
\begin{equation}
\Delta V_{g}\propto \Delta n_{2D}\propto T_{c}^{3/2},  \label{eq7}
\end{equation}
we obtain $z\overline{\nu }=2/3$ if $\Delta V_{g}$ or $\Delta n_{2D}$ are
taken as scaling argument $\delta $. To identify the correct scaling
argument we use the fact that $\delta \propto \Delta n_{2D}$ holds if $\left( 2+z\right) \overline{\nu }\geq 2$ \cite{fisher2}. To check this inequality, given $z\overline{\nu }$, we need an estimate of $z$.
For this purpose we invoke relation (\ref{eq4}), $R_{0c}-R_{0}\left(
T_{c}\right) \propto \xi _{0}^{-2}\left( T_{c}\right) $ ($z_{cl}=2$) which
diverges as $\xi _{0}(T_{c})\propto \xi \left( T=0\right) \propto \delta ^{-
\bar{\nu}}\propto T_{c}^{-1/z}$, so that the scaling relation
\begin{equation}
\Delta R_{0}\left( T_{c}\right) =R_{0c}-R_{0}\left( T_{c}\right) \propto \xi
_{0}^{-2}\left( T_{c}\right) \propto \delta ^{2\overline{\nu }}\propto
T_{c}^{2/z},  \label{eq8}
\end{equation}
holds. Fig. \ref{fig3} depicts the $T_{c}$ dependence of $\xi _{0}\left( T_{c}\right) \propto \left( R_{0c}-R_{0}\left(
T_{c}\right) \right) ^{-1/2}$ and $b$, which is related to the vortex energy
$E_{c}$ (Eq. (\ref{eq1a})). Approaching the 2D-QSI-transition we observe
that the data point to $\xi _{0}\left( T_{c}\right) \propto 1/T_{c}$,
yielding for $z$ the estimate $z\simeq 1$ so that $\bar{\nu}\simeq 2/3$ with
$z\bar{\nu}\simeq 2/3$. As these exponents satisfy the inequality $\left(
2+z\right) \overline{\nu }\geq 2$ \cite{fisher2} we identified the correct
scaling argument, $\delta \propto \Delta n_{2D}\propto \Delta V_{g}$. From
the gate voltage dependence of the normal state sheet resistance shown in
Fig. 2 it also follows that the normal state conductivity scales as $\Delta
\sigma _{\square }\left( T^{\ast }\right) =\sigma _{\square }\left( T^{\ast
}\right) -\sigma _{\square c}\left( T^{\ast }\right) \propto \Delta
V_{g}^{2/3}$. The 2D-QSI transition is then characterized by the scaling
relations
\begin{eqnarray}
T_{c} &\propto &\delta ^{2/3}\propto \Delta R\left( T^{\ast }\right) \propto
\Delta R_{0}^{1/2}\left( T_{c}\right) \propto \Delta V_{g}^{2/3}\propto
\Delta n_{2D}^{2/3}  \nonumber \\
&\propto &\Delta \sigma _{\square }\left( T^{\ast }\right) \propto \xi
_{0}^{-1}\left( T_{c}\right) ,  \label{eq9}
\end{eqnarray}
where $\Delta \sigma _{\square }\left( T^{\ast }\right) \propto \Delta
n_{2D}^{2/3}$\ reveals non-Drude behavior in the normal state. The product $z\overline{\nu }\simeq 2/3$ agrees with that
found in the electric field effect tuned 2D-QSI transition in amorphous
ultrathin bismuth films \cite{parendo} and the magnetic-field-induced 2D-QSI
transition in Nb$_{0.15}$Si$_{0.85}$ films \cite{aubin}. On the contrary it
differs from the value $z\overline{\nu }\simeq 1$ that has been found in
thin NdBa$_{2}$Cu$_{3}$O$_{7}$ films using the electric-field-effect
modulation of the transition temperature \cite{matthey}. In any case our
estimates, $z\simeq 1$ and $\bar{\nu}\simeq 2/3$ point \ to a 2D-QSI
transition which belongs to the 3D-xy universality class \cite{book}.
Fig. \ref{fig3} also depicts the $T_{c}$ dependence of $b$, which is related to the
vortex energy. Since $b$ tends to a constant in the limit $T_{c}\rightarrow 0
$, Eq. (\ref{eq1a}) implies $db/dT_{c}=0$ and therewith $E_{c}(T_{c})\propto
k_{B}T_{c}$, while the vortex core radius $\xi _{0}$ diverges as $\xi
_{0}\left( T_{c}\right) \propto 1/T_{c}$, in analogy to the behavior of
superfluid $^{4}$He films \cite{williams}. The 2D-QSI transition is then also characterized
by an infinite vortex core radius and vanishing vortex core energy. As $T_{c}
$ increases from the 2D-QSI transition, $\xi _{0}$ drops, while the vortex
core energy increases. Finally, after passing the maximum $T_{c}$ the vortex
core radius continues to decrease with reduced $T_{c}$ while b increases
further.
\begin{figure}[tbp]
\centering
\includegraphics[angle=0,width=8.6cm]{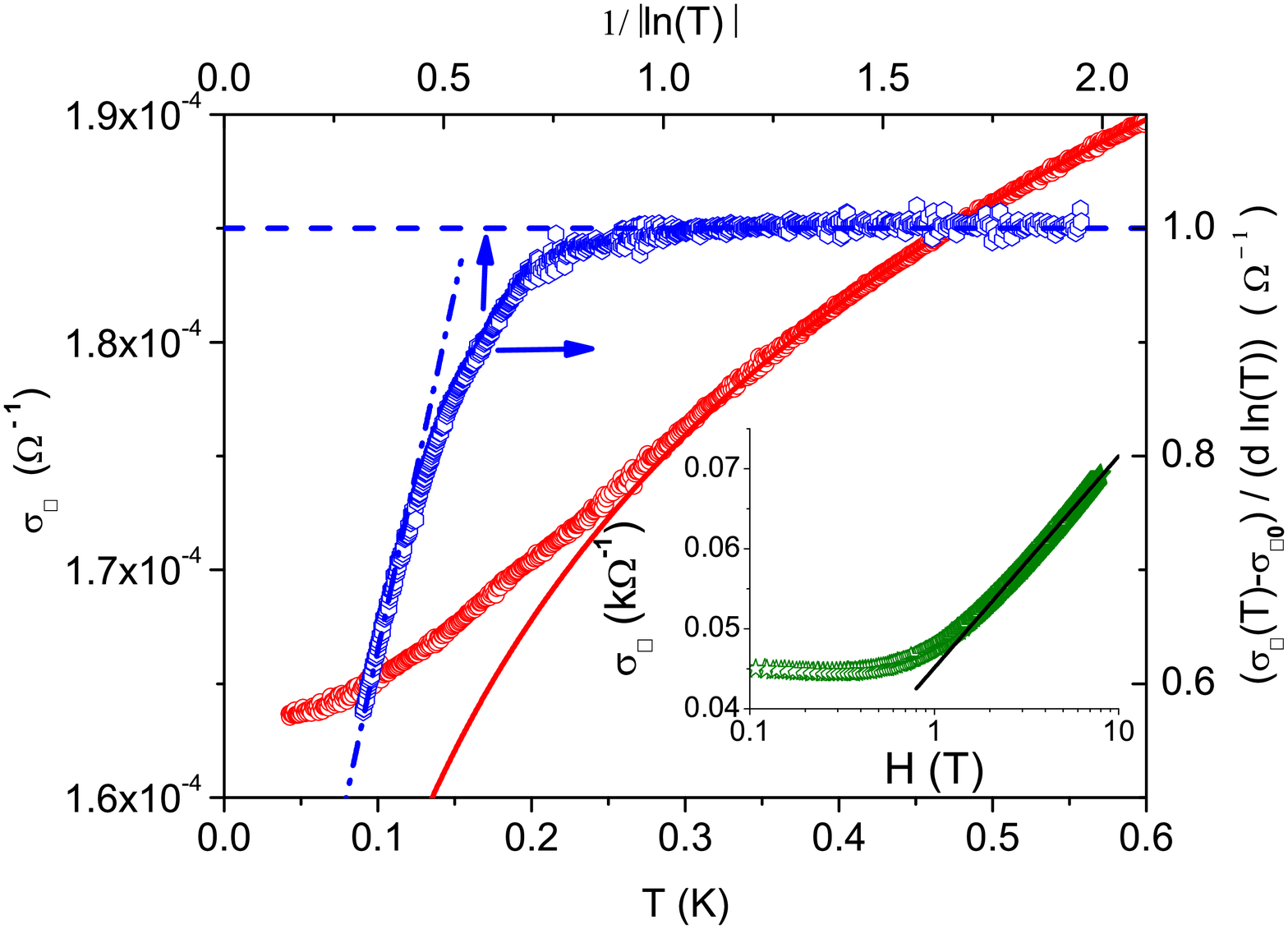}
\caption{$\sigma _{\square }$ \textit{vs}. $T$ and $\left( \sigma _{\square
}\left( T\right) -\sigma _{\square 0}\right) /\left( d\ln (T\right) )$
\textit{vs}. $1/\left\vert ln(T)\right\vert $ at $V_{g}=-240$ V from
Caviglia \textit{et al}. \cite{andrea}. The solid line is $\sigma _{\square
}\left( T\right) =\sigma _{\square 0}+d$ln$(T)$ with $\sigma _{\square
0}=1.2\cdot 10^{-4}\left( \Omega ^{-1}\right) $ and $d=1.2\cdot
10^{-5}\left( \Omega ^{-1}\right) $, the dashed one $\left( \sigma _{\square
}\left( T\right) -\sigma _{\square 0}\right) /\left( d\ln \left( T\right)
\right) =1$ and the dash-dot one $\left( \sigma _{\square }\left( T\right)
-\sigma _{\square 0}\right) /d=1.8\propto 1/L$. The inset shows the
magnetoconductivity $\sigma _{\square}$ \textit{vs}. $H$, applied
perpendicular to the interface, at $T=0.03$ K and $V_{g}=-300$ V, from
Caviglia \textit{et al}. \cite{andrea}. The solid line is $\sigma _{\square
}=4.51\cdot 10^{-2}+1.12\cdot 10^{-2}\ln \left( H\right) $ k$\Omega ^{-1}$.}
\label{fig4}
\end{figure}

Having presented the evidence for an electric field effect tuned BKT line
ending at a 2D-QSI-transition belonging to the 3D-xy universality class and
non-Drude behavior in the normal state an important issue remains, the
nature of the insulating phase. Fig. \ref{fig4} shows the temperature
dependence of $\sigma _{\square }$ for $V_{g}=-240$ V, which is rather deep
in the insulating phase \cite{andrea} . In analogy to Fig. \ref{fig1} we
observe a rounded transition. In the present case the intermediate regime is
compatible with weak localization, $\sigma _{\square }\left( T\right)
=\sigma _{\square 0}+d\ln (T)$ \cite{lee}. Given the evidence for the
limiting length $L$, which prevented the attainment of the superconducting
ground state, one suspects that it limits the growth of $\xi _{ld}\propto
R_{\square }=1/\sigma _{\square }$, the diverging length associated with
localization \cite{lee}, as well. In this case finite size scaling predicts
that $\sigma _{\square }(T)$ scales as $\left( \sigma _{\square }(T)-\sigma
_{\square c}\right) /\left( d\ln \left( T\right) \right) =g\left( y\right) $
with $y=\xi _{ld}/L\propto 1/\left( L\left\vert \ln \left( T\right)
\right\vert \right) $. $g(y)$ is the finite size scaling function and tends
to $1$ for $y<1$. In this case the approach to the insulating ground state
can be seen, while for $y>1$ the crossover to $g(y)\rightarrow y$ sets in
and $\sigma _{\square }(T)$ approaches the finite size dominated regime,
where $\sigma _{\square }(T)-\sigma _{\square 0}\propto 1/L$. A glance at
Fig. \ref{fig4} reveals that weak localization combined with the finite size
effect describes the data very well. In addition, below $V_{g}=$ $-300$ V $d$
is independent of $V_{g}$ and its value $d$ $=1.2\cdot 10^{-5}$ $\Omega ^{-1}
$ is close to $d_{e}=e^{2}/\left( \pi h\right) \simeq 1.23\cdot 10^{-5}$ $%
\Omega ^{-1}$, generically attributed to electron-electron interaction \cite
{blanter}. The resulting evidence for weak localization is further
substantiated by the observed negative magnetoresistance \cite{andrea} and
in particular by the high field behavior of the conductance depicted in the
inset of Fig. \ref{fig4}. There we observe consistency with the
characteristic ln$(H)$ high field behavior \cite{lee} and $d\sigma _{\square
}/d\ln \left( H\right) \simeq 1.12\cdot 10^{-5}\Omega ^{-1}$ is close to the
theoretical prediction $d\sigma _{\square }/d\ln \left( H\right)
=e^{2}/\left( \pi h\right) \simeq 1.23\cdot 10^{-5}\Omega ^{-1}$ \cite{altshuler}. As a result, the failure to observe the superconducting and insulating ground states directly, is attributable to a finite size effect,
preventing the respective diverging lengths to grow beyond $L$.
Nevertheless, the finite size scaling analysis and the high field
magnetoconductivity provide substantial evidence that these are the
appropriate ground states in the homogenous and infinite system.

 In summary, we have shown that the electrostatically tuned phase transition
line at the LaAlO$_{3}$/SrTiO$_{3}$ interface, observed by Caviglia \textit{
et al}. \cite{andrea}, is consistent with a BKT-line ending at a 2D-QSI
critical point \ with critical exponents $z\simeq 1$ and $\overline{\nu }%
\simeq 2/3$, so the universality class of the transition appears to be that
of the classical 3D-xy model. The normal state was shown to exhibit
non-Drude behavior. To identify the superconducting and insulating ground
states from the temperature dependence of the resistance we performed a
finite size scaling analysis because the growth of the finite temperature
correlation length and the diverging length associated with localization
turned out to be limited. Nevertheless, in the insulating state we observed
in both, the temperature and magnetic field dependence of the resistance, the
characteristic weak localization behavior. Furthermore, we explored the $
T_{c}$ dependence of the vortex core radius and the vortex energy. These
properties appear to be basic ingredients to unravel the nature of the
variation of $T_{c}$. Approaching the 2D-QSI transition the vortex energy
tends to zero while the vortex core radius and the localization length
diverge so the system is an insulator.

 This work was partially supported by the Swiss National Science Foundation through the National Center of Competence in
Research, and ``Materials with Novel Electronic Properties, MaNEP'' and Division II.

\end{document}